\documentstyle[12pt]{article}
\input psfig

\renewcommand{\baselinestretch}{1.2}
\newcommand{\nc}{\newcommand}

\nc{\eqr}[1]{(\ref{#1})}
\nc{\sref}[1]{\S\ref{#1}}
\nc{\tref}[1]{Table~\ref{#1}}
\nc{\fref}[1]{Figure~\ref{#1}}
\nc{\cref}[1]{Chapter~\ref{#1}}
\nc{\beq}{\begin{equation}}
\nc{\eeq}{\end{equation}}
\nc{\ben}{\begin{enumerate}}
\nc{\een}{\end{enumerate}}
\nc{\setall}{\setcounter{equation}{0}
        \setcounter{definition}{0}
        \setcounter{lemma}{0}
        \setcounter{convention}{0}
        \setcounter{conjecture}{0}
        \setcounter{theorem}{0}
        \setcounter{proposition}{0}
        \setcounter{property}{0}
        \setcounter{fact}{0}
        \setcounter{corollary}{0}}
\nc{\setequation}{\setcounter{equation}{0}}
\nc{\hs}[1]{\hspace{#1 mm}}

\def\sCC{{\kern 0.27em\vrule height1.45ex width0.03em depth0em
          \kern-0.30em\rm C}}
\def\C{{\mathchoice
  {\sCC}
  {\sCC}
  {\kern 0.225em \vrule height1.05ex width0.025em depth0em \kern-0.25em \rm C}
  {\kern 0.180em \vrule height0.78ex width0.02em depth0em \kern-0.2em \rm C}
        }}
\def\sHH{{\rm I\kern-.16em{}H}}
\def\H{{\mathchoice
  {\sHH}
  {\sHH}
  {\rm I\kern-.13em{}H}
  {\rm I\kern-.13em{}H} }}
\def\sNN{{\rm I\kern-.16em{}N}}
\def\N{{\mathchoice
  {\sNN}
  {\sNN}
  {\rm I\kern-.12em{}N}
  {\rm I\kern-.10em{}N} }}
\def\sPP{{\rm I\kern-.16em{}P}}
\def\P{{\mathchoice
  {\sPP}
  {\sPP}
  {\rm I\kern-.12em{}P}
  {\rm I\kern-.10em{}P} }}
\def\sQQ{{\kern 0.27em \vrule height1.45ex width0.03em depth0em
          \kern-0.30em \rm Q}}
\def\Q{{\mathchoice
        {\sQQ}
        {\sQQ}
  {\kern 0.225em \vrule height1.05ex width0.025em depth0em \kern-0.25em \rm Q}
  {\kern 0.180em \vrule height0.78ex width0.020em depth0em \kern-0.20em \rm Q}
        }}
\def\sRR{{\rm I\kern-0.16em{}R}}
\def\R{{\mathchoice
  {\sRR}
  {\sRR}
  {\rm I\kern-0.12em{}R}
  {\rm I\kern-0.10em{}R} }}
\def\sZZ{{\rm Z\kern-0.32em{}Z}}
\def\Z{{\mathchoice
  {\sZZ}
  {\sZZ} 
  {\rm Z\kern-0.3em{}Z}     
  {\rm Z\kern-0.25em{}Z} }}  
\def\ZZZ{{\rm Z\kern-0.24em{}Z}}
\def\sKK{{\rm I\kern-0.16em{}K}}
\def\K{{\mathchoice
  {\sKK}
  {\sKK}
  {\rm I\kern-0.12em{}K}
  {\rm I\kern-0.10em{}K} }}

\newtheorem{definition}{\bf DEFINITION}

\newtheorem{lemma}{\bf LEMMA}

\newtheorem{theorem}{\bf THEOREM}

\newtheorem{proposition}{\bf PROPOSITION}

\newtheorem{corollary}{\bf COROLLARY}

\newtheorem{observation}{\bf OBSERVATION}

\begin{document}
\renewcommand{\baselinestretch}{1}

\thispagestyle{empty}
{\flushright{\small MIT-CTP-2921\\ \small NSF-ITP-99-137\\hep-th/9911114\\}}

\vspace{.3in}
\begin{center}\LARGE {Some Remarks on the Finitude of Quiver Theories}
\end{center}

\vspace{.2in}
\begin{center}
{\large Yang-Hui He\\}
\normalsize{yhe@ctp.mit.edu\footnote{
Research supported in part
by the CTP and the LNS of MIT and the U.S. Department of Energy
under cooperative research agreement $\#$DE-FC02-94ER40818, the ITP of
UCSB under NSF grant PHY94-07194, as well as
the NSF Graduate Fellowship.}
\\}
\vspace{.2in} {\it Center for Theoretical Physics,\\ Massachusetts
Institute of Technology\\ Cambridge, MA 02139, U.S.A.\\}
\vspace{.1in} {\it and \\}
\vspace{.2in} {\it Institute for Theoretical Physics, \\
	UCSB, Santa Barbara, CA 93106, U.S.A.\footnote{
		Address until Nov. 21st, 1999.}\\}
\end{center}
\vspace{0.1in}

\begin{abstract}
D-brane probes, Hanany-Witten setups and geometrical engineering
stand as a trichotomy of the currently fashionable techniques of
constructing gauge theories from string theory. Meanwhile, asymptotic
freedom, finitude and IR freedom pose as a trichotomy of the
beta-function behaviour in quantum field theories.
Parallel thereto is a trichotomy in set theory of finite, tame and
wild representation types.
At the intersection of the above lies the theory of quivers.
We briefly review some of the terminology standard to the physics and to
the mathematics. Then we utilise certain results from graph theory and
axiomatic representation theory of path algebras to address physical
issues such as the implication of graph additivity to finiteness of
gauge theories, the impossibility of constructing completely IR free
string orbifold theories and the unclassifiability of ${\cal N}<2$
Yang-Mills theories in four dimensions.
\end{abstract}

\vspace{1.5in}

\pagebreak

\section{Introduction}
In a quantum field theory (QFT), it has been known since the 70's (q.v. e.g. 
\cite{Peskin}), that
the behaviour of physical quantities such as mass and coupling constant are
sensitive to the renormalisation and evolve according to momentum scale
as dictated by the so-called {\it renormalisation flows}. In particular,
the correlation (Green's) functions, which encode the physical 
information relevant to Feymann's perturbative analysis of the theory and
hence unaffected by such flows, obey the famous Callan-Symanzik
Equations.
These equations assert the existence two universal functions $\beta(\lambda)$
and $\gamma(\lambda)$
shifting according to the coupling and field renormalisation
in such a way so as to compensate for the renormalisation scale.

A class of QFT's has lately received much attention, particularly among
the string theorists. These are the so-named {\bf finite} theories, 
characterised by the vanishing of the $\beta$-functions. These theories are
extremely well-behaved and no divergences can be associated with the coupling
in the ultraviolet; they were thus once embraced as the solution to ultraviolet
infinities of QFT's. Four-dimensional finite theories are restricted to 
supersymmetric gauge theories (or Super-Yang-Mills, SYM's), of which divergence
cancelation is a general feature, and have a wealth of interesting structure.
${\cal N} = 4$ SYM theories have been shown to be finite to all orders
(Cf. e.g. \cite{West,N4}) whereas for ${\cal N}=2$, the Adler-Bardeen
Theorem guarantees that no higher than 1-loop corrections exist for the 
$\beta$-function \cite{N2}. Finally, for the unextended ${\cal N}= 1$ 
theories, the vanishing at 1-loop implies that for 2-loops \cite{N1}.

When a {\bf conformal} field theory (CFT) with vanishing
$\beta$-function also has the anomalous dimensions vanishing, the
theory is in fact a finite theory.
This class of theories is without divergence and scale -- and here we enter
the realm of string theory. Recently much attempts have been undertaken in the
construction of such theories as low-energy limits of the world-volume theories
of D-brane probes on spacetime singularities \cite{Orb1,Orb2,KS,LNV,Han-He} or 
of brane setups of the Hanany-Witten type \cite{HW,Han-S,Han-Zaf,Han-Ura}.
The construction of these theories not only supplies an excellent check for
string theoretic techniques but also, vice versa, facilitate the
incorporation of the Standard Model into string unifications.
These finite (super-)conformal theories in four dimensions
still remain a topic of fervent pursuit.

Almost exactly concurrent with these advances in physics was a host of 
activities in mathematics. Inspired by problems in linear representations
of partially ordered sets over a field \cite{Simson,Gabriel1,Gabriel2,Sets,Dlab},
elegant and graphical methods have been
developed in attacking standing problems in algebra and combinatorics such
as the classification of representation types and indecomposables of 
finite-dimensional algebras.

In 1972, P. Gabriel introduced the concept of a ``K\"ocher''
in \cite{Gabriel1}. This is what is known to our standard parlance today 
as a ``Quiver.'' What entailed
was a plethora of exciting and fruitful research in graph theory, axiomatic
set theory, linear algebra and category theory, among many other branches. 
In particular one result that has spurned interest is the great limitation
imposed on the shapes of the quivers once the concept of {\bf finite representation
type} has been introduced.

It may at first glance seem to the reader that these two disparate directions of
research in contemporary physics and mathematics may never share conjugal harmony.
However, following the works of
\cite{Orb2,KS,LNV} those amusing quiver diagrams have 
surprisingly - or perhaps
not too much so, considering how that illustrious field of String Theory has
of late brought such enlightenment upon physics from seemingly most 
esoteric mathematics - taken a slight excursion from the reveries of the
abstract, and manifested themselves in SYM theories emerging
from D-branes probing orbifolds.
The gauge fields and matter content of
the said theories are conveniently encoded into quivers and further
elaborations upon relations to beyond orbifold theories have been suggested in
\cite{Han-He,He-Song}.

It is therefore natural, for one to pause and step back awhile, and regard the
string orbifold theory from the perspective of a mathematician, and the quivers,
from that of a physicist. However, due to his inexpertise in both, the author
could call himself neither. Therefore we are compelled to peep at the
two fields as outsiders, and from afar attempt to make some observations
on similarities, obtain some vague notions of the beauty, and speculate upon
some underlying principles. This is then the purpose of this note:
to perceive, with a distant and weak eye; to inform, with a remote and feeble voice.

The organisation of the paper is as follows. Though the main results
are given in \S 4, we begin with some preliminaries from
contemporary techniques in string theory on constructing four dimensional
super-Yang-Mills, focusing on what each interprets finitude to mean: 
\sref{ss:probe} on D-brane probes on orbifold singularities,
\sref{ss:HW} on Hanany-Witten setups and \sref{ss:geo} on geometrical
engineering. Then we move to the other direction and give preliminaries in
the mathematics, introducing quiver graphs and path 
algebras in \sref{ss:quiver},
classification of representation types in \sref{ss:type} and
how the latter imposes constraints on the former in \sref{ss:theorem}.
The physicist may thus liberally neglect \S 2 and the mathematician, 
\S 3. Finally in \sref{s:marriage} we shall see how those beautiful theorems
in graph theory and axiomatic set theory may be used to give surprising
results in constructing gauge theories from string theory.

\section*{Nomenclature}
Unless the contrary is stated, we shall throughout this paper adhere to 
the convention that $k$ is a field of characteristic zero (and hence infinite),
that $Q$ denotes a quiver and $kQ$, the path algebra over the field $k$ associated
thereto, that rep$(X)$ refers to the representation of the object $X$, and that
irrep($\Gamma$) is the set of irreducible representations of the group $\Gamma$.
Moreover, {\sf San serif} type setting will be reserved for categories,
calligraphic ${\cal N}$ is used to denote the number of supersymmetries and
$\widehat{~~}$, to distinguish the Affine Lie Algebras or Dynkin graphs.

\section{Preliminaries from the Physics}
The Callan-Symanzik equation of a QFT dictates the behaviour, under
the renormalisation 
group flow, of the $n$-point correlator $G^{(n)}(\{\phi(x_i)\};M,\lambda)$ 
for the quantum fields $\phi(x)$,
according to the renormalisation of the coupling $\lambda$ and momentum
scale $M$ (see e.g. \cite{Peskin}, whose conventions we shall adopt):
\[
\left[ M \frac{\partial}{\partial M} + \beta(\lambda) \frac{\partial}{\partial \lambda}
+ n \gamma(\lambda) \right] G^{(n)}(\{\phi(x_i)\}; M,\lambda) = 0.
\]
The two universal dimensionless functions $\beta$ and $\gamma$ are known respectively
as the {\bf $\beta$-function} and the {\bf anomalous dimension}. They determine how the
shifts $\lambda \rightarrow \lambda + \delta \lambda$ in the coupling constant and
$\phi \rightarrow (1 + \delta \eta) \phi$ in the wave function compensate for the
shift in the renormalisation scale $M$:
\[
\beta(\lambda) := M \frac{\delta \lambda}{\delta M}~~~~~~
\gamma(\lambda) := -M \frac{\delta \eta}{\delta M}.
\]
Three behaviours are possible in the region of small $\lambda$:
(1) $\beta(\lambda) > 0$;
(2) $\beta(\lambda) < 0$; and
(3) $\beta(\lambda) = 0$.
The first has good IR behaviour and admits valid Feynmann perturbation at large-distance,
and the second possesses good perturbative behaviour at UV limits and are
asymptotically free. The third possibility is where the coupling constants do not
flow at all and the renormalised coupling is always equal to the bare coupling.
The only possible divergences in these theories are associated with field-rescaling
which cancel automatically in physical $S$-matrix computations. It seems that to
arrive at these well-tamed theories, some supersymmetry (SUSY) is needed so as to induce the
cancelation of boson-fermion loop effects\footnote{Proposals for non-supersymmetric
finite theories in four dimensions have been recently made in \cite{KS,LNV,FV,Su4}; to their
techniques we shall later turn briefly.}.
These theories are known as the {\bf finite theories} in QFT.

Of particular importance are the finite theories 
that arise from {\bf conformal} field theories which generically have
in addition to the vanishing $\beta$-functions, also zero anomalous
dimensions.
Often this subclass belongs to a continuous manifold of scale invariant
theories and is characterised by the existence of exactly marginal operators and whence
dimensionless coupling constants, the
set of mappings among which constitutes the {\it duality group} \`a la Mantonen-Olive
of ${\cal N} = 4$ SYM, a hotly pursued topic.

A remarkable phenomenon is that if there is a choice of coupling constants such that
all $\beta$-functions as well as the anomalous dimensions (which themselves do vanish
at leading order if the manifold of fixed points include the free theory) vanish at first 
order then the theory is finite to all orders (Cf. references in \cite{Han-S}).
A host of finite theories arise as low energy effective theories of String Theory. It will
be under this light that our discussions proceed. There are three contemporary methods of
constructing (finite, super) gauge theories: (1) geometrical engineering; (2)
D-branes probing singularities and (3) Hanany-Witten brane setups. Discussions on the 
equivalence among and extensive reviews for them have been in wide circulation 
(q.v. e.g. \cite{Methods,ZD}). Therefore we shall not delve too far into their account;
we shall recollect from them what each interprets {\it finitude} to mean.

\subsection{D-brane Probes on Orbifolds} \label{ss:probe}
When placing $n$ D3-branes on a space-time orbifold singularity $\C^m/\Gamma$, 
out of the parent
${\cal N} = 4$ $SU(n)$ SYM one can fabricate a $\prod\limits_{i} U(N_i)$ 
gauge theory with irrep$(\Gamma) := \{{\bf r}_i \}$ 
and $\sum\limits_{i} N_i \dim{\bf r}_i = n$ \cite{LNV}. The resulting SUSY in
the four-dimensional worldvolume is ${\cal N} = 2$ if the orbifold is
$\C^2 / \{ \Gamma \subset SU(2) \}$ as studied in \cite{Orb2},
${\cal N} = 1$ if $\C^3 / \{ \Gamma \subset SU(3) \}$ as in \cite{Han-He}
and non-SUSY if $\C^3 / \{ \Gamma \subset SU(4) \}$ as in \cite{Su4}.
The subsequent matter fields are
$a_{ij}^{\bf{4}}$ Weyl fermions $\Psi _{f_{ij}=1,...,a_{ij}^{\bf{4}}}^{ij}$ 
and $a_{ij}^{\bf 6}$ scalars $\Phi _{f_{ij}}^{ij}$ with $i,j = 1,...,n$ and
$a_{ij}^{\cal R}$ defined by
\beq
{\cal R}\otimes {\bf r}_i=\bigoplus\limits_{j}a_{ij}^{\cal R} {\bf r}_j
\label{aij}
\eeq
respectively for ${\cal R} = 4,6$.
It is upon these matrices $a_{ij}$, which we call {\bf bifundamental
matter matrices} that we shall dwell. They dictate how many matter fields
transform under the $(N_i,\bar{N}_j)$ of the product gauge group.
It was originally pointed out in \cite{Orb2} that one can encode this information
in {\bf quiver diagrams} where one indexes the vector multiplets (gauge) by nodes
and hypermultiplets (matter) by links in a (finite) graph so that the bifundamental
matter matrix defines the (possibly oriented) adjacency matrix for this graph.
In other words, one draws $a_{mn}$ number of arrows from node $m$ to $n$.
Therefore to each vertex $i$ is associated a vector space $V_i$ and a semisimple 
component $SU(N_i)$ of the gauge group acting on $V_i$. Moreover 
an oriented link from $V_1$ to $V_2$ represents a complex field transforming under
hom($V_1,V_2$). We shall see in section \sref{ss:quiver} what all this means.

When we take the dimension of both sides of (\ref{aij}), we obtain the matrix
equation 
\beq
\label{dimaij}
\dim({\cal R}) r_i = a_{ij}^{\cal R} r_j
\eeq
where $r_i := \dim{\bf r}_i$.
As discussed in \cite{LNV,Han-He}, the remaining SUSY must be in the 
commutant of $\Gamma$ in the $SU(4)$ R-symmetry of the parent
${\cal N} = 4$ theory. In the case of ${\cal N} = 2$ this means that 
$4 = 1 + 1 + 2$ and by SUSY, $6 = 1 + 1 + 2 + 2$ where the 1 is the principal
(trivial) irrep and 2, a two-dimensional irrep. Therefore due to the additivity and
orthogonality of group characters, it was thus pointed out ({\it cit. ibid.}) that
one only needs to investigate the fermion matrix $a_{ij}^4$, which is actually
reduced to $2 \delta_{ij} + a_{ij}^2$. Similarly for ${\cal N} = 1$, we have
$\delta_{ij} + a_{ij}^3$. It was subsequently shown that
(\ref{dimaij}) necessitates the vanishing of the $\beta$-function
to one loop. Summarising these points, we state the condition for {\it finitude}
from the orbifold perspective:
\beq
\label{orb_cond}
\begin{array}{c|c}
$SUSY$		&	$Finitude$ \\ \hline
{\cal N} = 2	&	2 r_i = a_{ij}^2 r_j \\
{\cal N} = 1	&	3 r_i = a_{ij}^3 r_j \\
{\cal N} = 0	&	4 r_i = a_{ij}^4 r_j \\
\end{array}
\eeq
In fact it was shown in \cite{LNV,Mirror}, that the 1-loop $\beta$-function
is proportional to $d r_i - a_{ij}^d r_j$ for $d=4 - {\cal N}$ whereby
the vanishing thereof signifies finitude, exceeding zero signifies asymptotical freedom
and IR free otherwise\footnote{As a cautionary note, these conditions
	are necessary but may not be sufficient. In the cases
	of ${\cal N}<2$, one needs to check the superpotential.
	However, throughout the paper we shall focus on the necessity
	of these conditions.}.
We shall call this expression $d \delta_{ij} - a_{ij}^d$
the {\bf discriminant function} since its relation with respective to zero
(once dotted with the vector of labels) discriminates the behaviour of the
QFT. This point shall arise once again in \sref{s:marriage}.

\subsection{Hanany-Witten} \label{ss:HW}
In brane configurations of the Hanany-Witten type \cite{HW}, D-branes
are stretched between sets of NS-branes, the presence of which break the
SUSY afforded by the 32 supercharges of the type II theory. In particular,
parallel sets of NS-branes break one-half SUSY, giving rise to ${\cal N} = 2$ in
four dimensions \cite{HW} whereas rotated NS-branes \cite{Erlich} or grids of NS-branes
(the so-called Brane Box Models) \cite{Han-S,Han-Zaf,Han-Ura} break one further half
SUSY and gives ${\cal N}=1$ in four dimensions.

The Brane Box Models (BBM) (and possible extensions to brane cubes) provide an intuitive
and visual realisation of SYM. They generically give rise to ${\cal N} = 1$, with
${\cal N} = 2$ as a degenerate case. Effectively, the D-branes placed in the
boxes of NS-branes furnish a geometrical way to encode the representation properties
of the finite group $\Gamma$ discussed in \sref{ss:probe}. The bi-fundamentals,
and hence the quiver diagram, are constructed from oriented open strings connecting
the D-branes according to the rule given in \cite{Han-Zaf}:
\[
{\bf 3} \otimes r_i = \bigoplus\limits_{j\in {\tiny \begin{array}{c}
{\rm N,~E,~SW}\\ {\rm Neighbours} \end{array}}} r_j.
\]
This is of course (\ref{aij}) in a different guise and we clearly see the equivalence
between this and the orbifold methods of \sref{ss:probe}.

Now in \cite{HW}, for the classical setup of stretching a D-brane between two NS-branes,
the asymptotic bending of the NS-brane controls the evolution of the gauge coupling
(since the inverse of which is dictated by the distance between the NS-branes).
Whence NS-branes bending towards each other gives an IR free theory (case (1) defined
above for the $\beta$-functions), while bending away give an UV free (case (2))
theory. No bending thus indicates the non-evolution of the $\beta$-function and
thus finiteness; this is obviously true for any brane configurations, intervals,
boxes or cubes. We quote \cite{Han-S} verbatim on this issue:
{\it Given a brane configuration which has no bending, the corresponding
field theory which is read off from the brane configuration by using the rules of
\cite{Han-Zaf} is a finite theory.}

Discussions on bending have been treated in \cite{LR,Randall} while works
towards the establishment of the complete correspondence between Hanany-Witten
methods and orbifold probes (to beyond the Abelian case) are well under
way \cite{ZD}. Under this light, we would like to lend this opportunity to point
out that the anomaly cancelation equations (2-4) of \cite{LR} which discusses the 
implication of tadpole-cancelation to BBM in excellent detail, are precisely
in accordance with (\ref{aij}). In particular, what they referred as the Fourier
transform to extract the rank matrix for the $\Z_k \times \Z_{k'}$ BBM is
precisely the orthogonality relations for finite group characters (which in the
case of the Abelian groups conveniently reduce to roots of unity and hence
Fourier series). The generalisation of these equations for non-Abelian groups
should be immediate. We see indeed that there is a close intimacy between
the techniques of the current subsection with \sref{ss:probe}; let us now move
to a slightly different setting.

\subsection{Geometrical Engineering} \label{ss:geo}
On compactifying Type IIA string theory on a non-compact Calabi-Yau 
threefold, we can geometrically engineer \cite{GeoEng,GeoMat,Mirror}
an ${\cal N}=2$ SYM.
More specifically when we compactify Type IIA on a K3 surface,
locally modeled by an ALE singularity, we arrive at an ${\cal N}=2$
SYM in 6 dimensions with gauge group $ADE$ depending on the singularity
about which D2-branes wrap in the zero-volume limit. However if we were
to further compactify on $T^2$, we would arrive at an ${\cal N} = 4$ SYM
in 4 dimensions. In order to kill the extraneous scalars we require
a 2-fold without cycles, namely {\bf P}$^1$, or the 2-sphere.
Therefore we are effectively compactifying our original 10 dimensional theory
on a (non-compact) Calabi-Yau threefold which is an ALE (K3) fibration
over {\bf P}$^1$, obtaining a pure ${\cal N}=2$ SYM in 4 dimensions with
coupling $\frac{1}{g^2}$ equaling to the volume of the base {\bf P}$^1$.

To incorporate matter \cite{GeoMat,Mirror} we let an $A_{n-1}$ ALE
fibre collide with an $A_{m-1}$ one to result in an $A_{m+n-1}$
singularity; this corresponds
to a Higgsing of $SU(m+n) \rightarrow SU(m) \times SU(n)$, giving
rise to a bi-fundamental matter $(n,\bar{m})$. Of course, by
colliding the $A$ singular fibres appropriately (i.e., in accordance with
Dynkin diagrams) this above idea can easily be generalised to fabricate
generic product $SU$ gauge groups.
Thus as opposed to \sref{ss:probe} where bi-fundamentals 
(and hence the quiver diagram) arise from linear maps between
irreducible modules of finite group representations, or \sref{ss:HW}
where they arise from open strings linking D-branes, in the context of
geometrical engineering, they originate from colliding fibres of the
Calabi-Yau.

The properties of the $\beta$-function from this geometrical perspective
were also investigated in \cite{Mirror}. The remarkable fact, using the
Perron-Frobenius Theorem, is that the {\it possible} resulting SYM is highly
restricted. The essential classification is that {\it if the ${\cal
N}=2$ $\beta$-
function vanishes (and hence a {\rm finite} theory), then the
quiver diagram encoding the bi-fundamentals must be the affine 
$\widehat{ADE}$ Dynkin Diagrams} and when it is less than zero (and
thus an asymptotically free theory), the quiver must be the ordinary
$ADE$. We shall see later how one may graphically arrive at these results.

Having thus reviewed the contemporary trichotomy of the methods of
constructing SYM from string theory fashionable of late,
with special emphasis on what the word {\it finitude} means in each,
we are obliged, as prompted by the desire to unify, to ask ourselves
whether we could study these techniques axiomatically. After all, the
quiver diagram does manifest under all these circumstances. And it is
these quivers, as viewed by a graph or representation theorist, that
we discuss next.

\section{Preliminaries from the Mathematics}
We now formally study what a quiver is in a mathematical sense.
There are various approaches one could take, depending on whether one's
interest lies in category theory or in algebra. We shall commence with
P. Gabriel's definition, which was the genesis of the excitement which ensued.
Then we shall introduce the concept of path algebras and representation types
as well as a host of theorems that limit the shapes of quivers depending on
those type. As far as convention and nomenclature are concerned, \sref{ss:quiver} and
\sref{ss:type} will largely follow \cite{Rep,Mod,Simson}.

\subsection{Quivers and Path Algebras} \label{ss:quiver}
In his two monumental papers \cite{Gabriel1,Gabriel2},
Gabriel introduced the following concept:
\begin{definition}
A {\bf quiver} is a pair $Q = (Q_0,Q_1)$, where $Q_0$ is a
set of vertices and $Q_1$, a set of arrows such that each element $\alpha \in Q_1$
has a beginning $s(\alpha)$ and an end $e(\alpha)$ which are vertices, i.e.,
$\{s(\alpha) \in Q_0\} \stackrel{\alpha}{\rightarrow} \{e(\alpha) \in Q_0$\}.
\end{definition}
In other words a quiver is a (generically) {\it directed graph}, 
possibly with multiple arrows and loops. 
We shall often denote a member $\gamma$ of $Q_1$ by the beginning and
ending vertices, as in $x \stackrel{\gamma}{\rightarrow} y$.

Given such a graph, we can generalise $Q_{0,1}$ by defining a {\bf path of length $m$} 
to be the formal composition
$\gamma = \gamma_1 \gamma_2 \dots \gamma_m := 
(i_0 \stackrel{\gamma_1}{\rightarrow} i_1 \dots \stackrel{\gamma_m}{\rightarrow} i_m)$
with $\gamma_j \in Q_1$ and $i_j \in Q_0$ such that $i_0 = s(\gamma_1)$ and 
$i_t = s(\gamma_{t-1}) = e(\gamma_t)$ for $t = 1,...,m$. This is to say that we
follow the arrows and trace through $m$ nodes. Subsequently we 
let $Q_m$ be the set of all paths of length $m$ and for the identity define,
for each node $x$, a trivial path of length zero, $e_x$, starting and ending at $x$.
This allows us to associate $Q_0 \sim \{e_x\}_{x \in Q_0}$ and 
$(i \stackrel{\alpha}{\rightarrow} j) \sim e_i \alpha = \alpha e_j$.
Now $Q_m$ is defined for all non-negative $m$, whereby giving a gradation in $Q$.

Objects\footnote{We could take this word literally and indeed we shall later
briefly define the objects in a Quiver Category.} may be assigned to the
nodes and edges of the quiver so as to make its conception more concrete. This is
done so in two closely-related ways:
\ben
\item By the {\bf representation of a quiver}, rep$(Q)$, we mean to 
	associate to each vertex $x \in Q_0$ of $Q$, a vector space $V_x$ 
	and to each arrow $x \rightarrow y$,
	a linear transformation between the corresponding vector spaces 
	$V_x \rightarrow V_y$.
\item Given a field $k$ and a quiver $Q$, a {\bf path algebra} $kQ$ is an algebra which 
	as a vector space over $k$ has its basis prescribed by the paths in $Q$.
\een

There is a 1-1 correspondence between $kQ$-modules and rep$(Q)$.
Given rep$(Q) = \{V_{x\in Q_0},(x \rightarrow y) \in Q_1 \}$, the associated
$kQ$ module is $\bigoplus\limits_x V_x$ whose basis is the set of paths $Q_m$.
Conversely, given a $kQ$-module $V$, we define $V_x = e_x V$ and the arrows to be
prescribed by the basis element $u$ such that $u \sim e_y u = u e_x$ whereby
making $u$ a map from $V_x$ to $V_y$.

On an algebraic level, due to the gradation of the quiver $Q$ by $Q_m$, the path
algebra is furnished by

\beq
\label{pathalg}
kQ := \bigoplus\limits_m kQ_m
~~~{\rm with}~~~
kQ_m := \bigoplus\limits_{\gamma \in Q_m} \gamma k
\eeq

As a $k$-algebra, the addition and multiplication axioms of $kQ$ are as
follows: given
$a = \sum\limits_{\alpha \in Q_m;~a_\alpha \in k} \alpha a_\alpha$
and
$b = \sum\limits_{\beta \in Q_n;~a_\beta \in k} \beta a_\beta$
as two elements in $kQ$,
$a+b = \sum\limits_\alpha \alpha (a_\alpha + b_\alpha)$ and
$a \cdot b = \sum\limits_{\alpha,\beta} \alpha \beta a_\alpha b_\beta$
with $\alpha \beta$ being the joining of paths (if the endpoint of
one is the beginning of another, otherwise it is defined to be 0).

\begin{figure}
\centerline{\psfig{figure=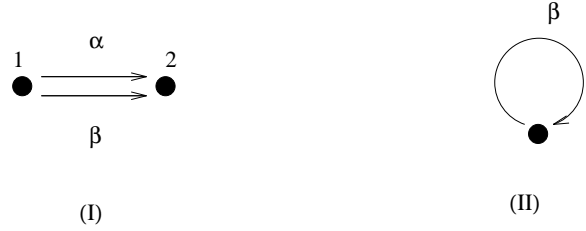,width=3.0in}}
\caption{Two examples of quivers with nodes and edges labeled.}
\label{f:ex}
\end{figure}

This correspondence between path algebras and quiver representations
gives us the flexibility of freely translating between the two, an advantage
we shall later graciously take.
As illustrative examples of concepts thus far introduced, we have
drawn two quivers in \fref{f:ex}.
In example (I), $Q_0 = \{1,2\}, Q_1 = \{\alpha,\beta\}$ and
$Q_{m > 1} = \{\}$. The path algebra is then the so-called Kronecker
Algebra: 
\[
kQ = e_1 k \oplus e_2 k \oplus \alpha k \oplus \beta k
= \left[ \begin{array}{cc} k & k^2 \\ 0 & k \\
	\end{array} \right].
\]
On the other hand, for example (II), 
$Q_{m \in \{0,1,2,...\}} = \{\beta^m\}$ and the
path algebra becomes $\bigoplus\limits_m \beta^m k = k[\beta]$,
the infinite dimensional free algebra of polynomials of one
variable over $k$.

In general, $kQ$ is finitely generated if there exists a finite
number of vertices and arrows in $Q$ and $kQ$ is finite-dimensional
if there does not exist any oriented cycles in $Q$.

\begin{figure}
\centerline{\psfig{figure=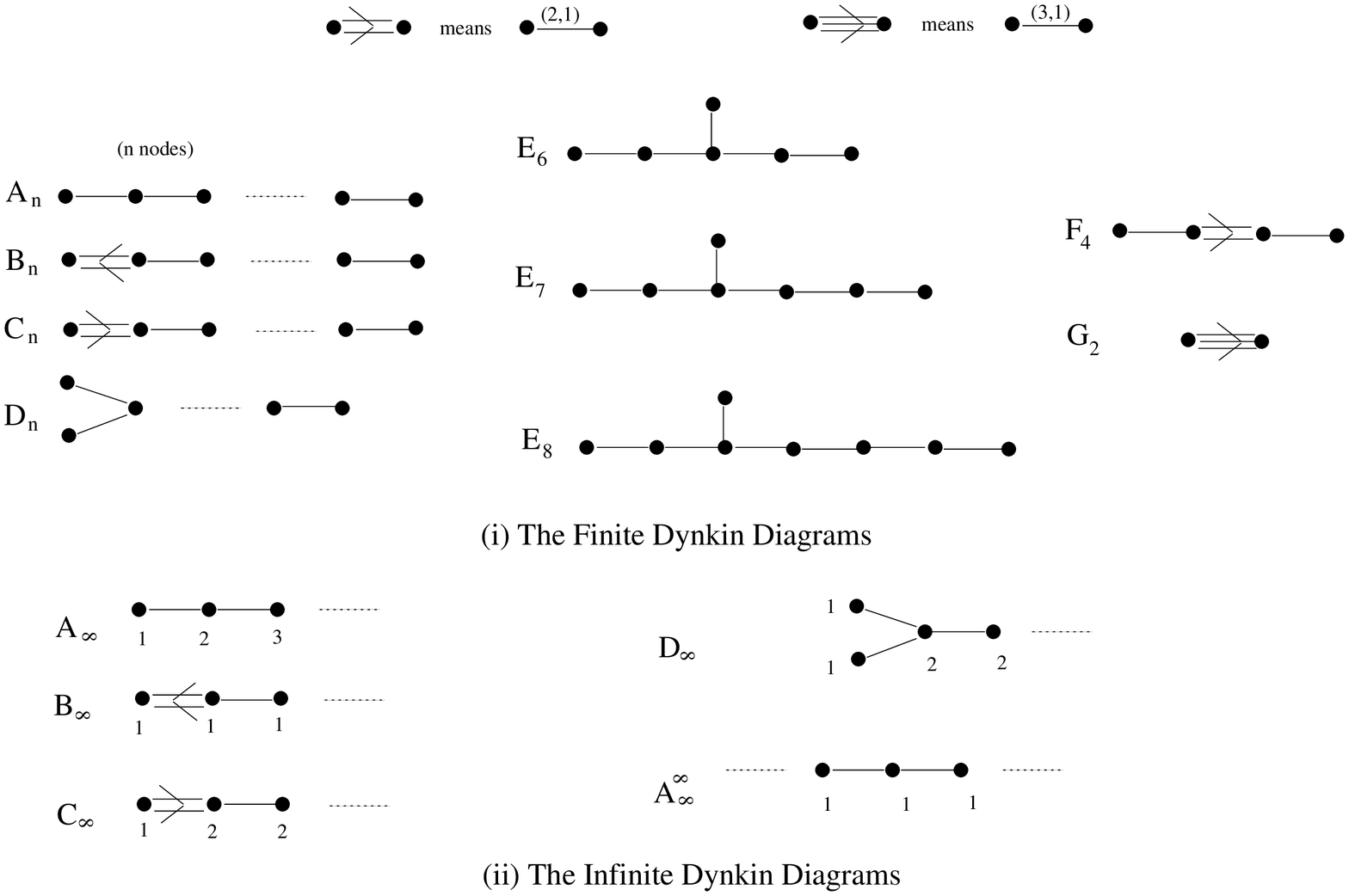,width=6.0in}}
\caption{The Finite and Infinite Dynkin Diagrams as labeled quivers.
	The finite cases are the well-known Dynkin-Coxeter graphs
	in Lie Algebras (from Chapter 4 of \cite{Rep}).}
\label{f:dynkin}
\end{figure}

\begin{figure}
\centerline{\psfig{figure=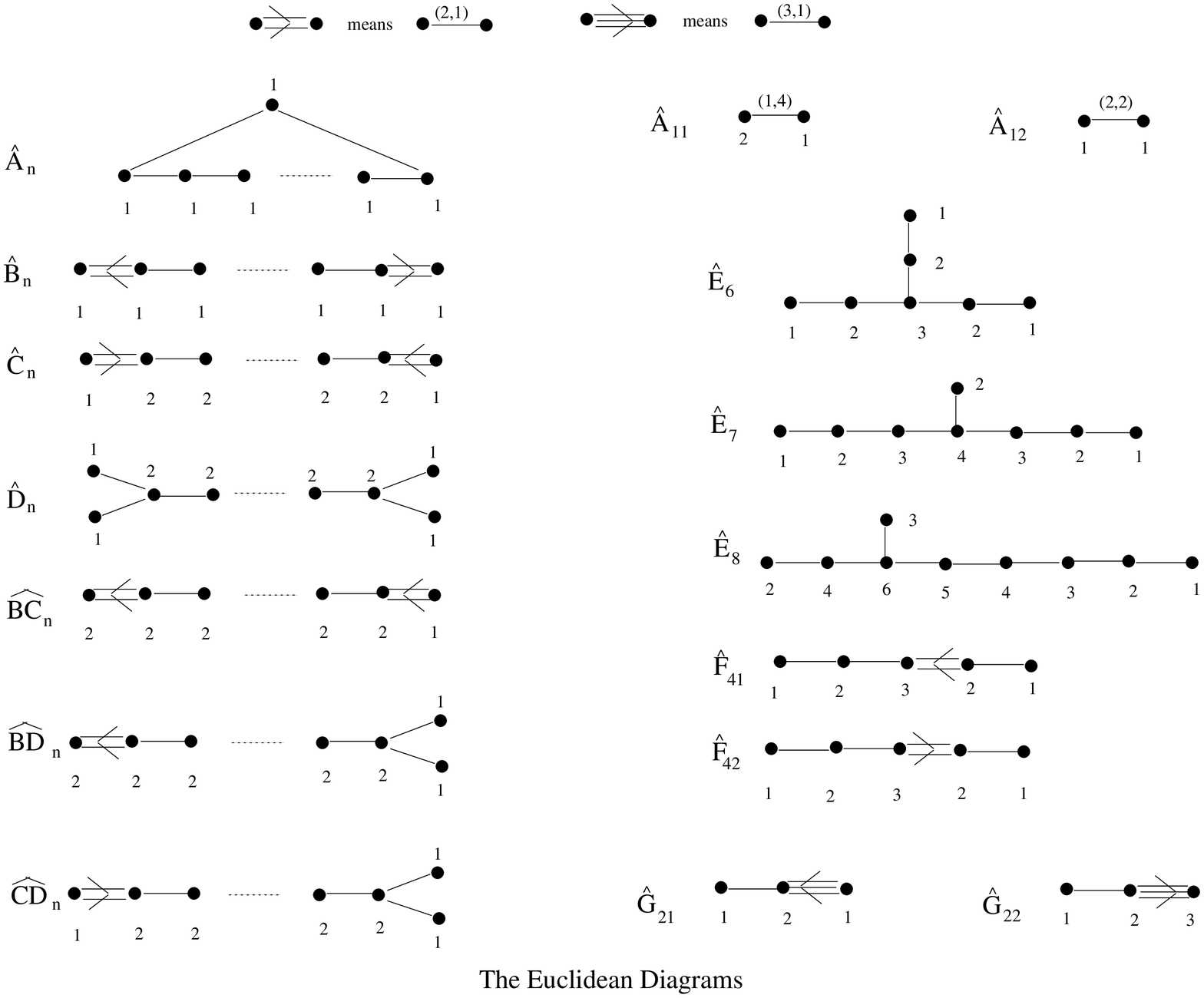,width=6.0in}}
\caption{The Euclidean Diagrams as labeled quivers; we recognise that
	this list contains the so-called Affine Dynkin Diagrams
	(from Chapter 4 of \cite{Rep}).}
\label{f:euclid}
\end{figure}

To specify the quiver even further one could introduce labeling
schemes for the nodes and edges; to do so we need a slight excursion
to clarify some standard terminology from graph theory.
\begin{definition} The following are common categorisations of graphs:
\begin{itemize}
\item A {\bf labeled graph} is a graph which has, for each of its edge
	$(i \stackrel{\gamma}{\rightarrow} j)$, a pair of positive
	integers $(a_{ij}^\gamma, a_{ji}^\gamma)$ associated thereto;
\item A {\bf valued graph} is a labeled graph for which there exists a
	positive integer $f_i$ for each node $i$, such that 
	$a_{ij}^\gamma f_j = a_{ji}^\gamma f_i$ for each arrow\footnote{
	Thus a labeled graph without any cycles is always a valued graph since
	we have enough degrees of freedom to solve for a consistent set of
	$f_i$ whereas cycles would introduce extra constraints.
	(Of course there is no implicit summation assumed in the equation.)}.
\item A {\bf modulation} of a valued graph consists of an assignment of
	a field $k_i$ to each node $i$, and a $k_i$-$k_j$
	bi-module $M_{ij}^\gamma$ to each arrow
	$(i \stackrel{\gamma}{\rightarrow} j)$ satisfying
	\ben
	\def\theenumi{\alph{enumi}}\def\labelenumi{(\theenumi)}
		\item $M_{ij}^\gamma \cong \hom_{k_i}(M_{ij}^\gamma,k_i)
			\cong \hom_{k_j}(M_{ij}^\gamma,k_j)$;
		\item $\dim_{k_i}(M_{ij}^\gamma) = a_{ij}^\gamma.$
	\een
\item A {\bf modulated quiver} is a valued graph with
	a modulation (and orientation).
\end{itemize}
\end{definition}
We shall further adopt the convention that we omit the label to edges
if it is $(1,1)$. We note that of course according to this labeling,
the matrices $a_{ij}$ are almost what we call {\bf adjacency matrices}.
In the case of unoriented single-valence edges between say nodes $i$ and
$j$, the adjacency matrix has $a_{ij} = a_{ji} = 1$, precisely the
label $(1,1)$. However, directed edges, as in \fref{f:dynkin}
and \fref{f:euclid}, are slightly more involved. This is
exemplified by $\bullet \Rightarrow \bullet$
which has the label $(2,1)$ whereas the conventional adjacency
matrix would have the entries $a_{ij} = 2$ and $a_{ji} = 0$. Such
a labeling scheme is of course so as to be consistent with the
entries of the Dynkin-Cartan Matrices of the semi-simple Lie Algebras.
To this subtlety we shall later turn.

The canonical examples of labeled (some of them are valued) graphs 
are what are known as the {\bf Dynkin} and {\bf Euclidean} graphs. 
The Dynkin graphs are further subdivided into the
finite and the infinite; the former are simply the Dynkin-Coxeter
Diagrams well-known in Lie Algebras while the latter are analogues
thereof but with infinite number of labeled nodes (note that 
the nodes are not labeled so as to make them valued graphs;
we shall shortly see what those numbers signify.)
The Euclidean graphs are the so-called Affine Coxeter-Dynkin Diagrams
(of the affine extensions of the semi-simple Lie algebras) but
with their multiple edges differentiated by oriented labeling schemes.
These diagrams are shown in \fref{f:dynkin} and \fref{f:euclid}.


How are these the {\it canonical} examples? We shall see the reason in 
\sref{ss:theorem}
why they are ubiquitous and atomic, constituting, when certain finiteness
conditions are imposed, the only elemental quivers.
Before doing so however, we need some facts from representation theory of
algebras; upon these we dwell next.

\subsection{Representation Type of Algebras} \label{ss:type}
Henceforth we restrict ourselves to infinite fields, as some of the
upcoming definitions make no sense over finite fields. This is of no loss 
of generality because in physics we are usually concerned with the field $\C$.
When given an algebra, we know its quintessential properties once we determine
its decomposables (or equivalently the irreducibles of the associated module).
Therefore classifying the behaviour of the indecomposables is the main goal
of classifying {\bf representation} types of the algebras.

The essential idea is that an algebra is of finite type if there are only
finitely many indecomposables; otherwise it is of infinite type.
Of the infinite type, there is one well-behaved subcategory, namely the
algebras of tame representation type, which has its indecomposables of
each dimension coming in finitely many one-parameter families with only
finitely many exceptions. Tameness in some sense still suggests
classifiability of the infinite indecomposables. On the other hand, an
algebra of wild type includes the free algebra on two variables, $k[X,Y]$,
(the path algebra of \fref{f:ex} (II), but with two self-adjoining arrows),
which indicates representations of arbitrary finite dimensional algebras,
and hence unclassifiability\footnote{
For precise statements of the unclassifiability of modules of
two-variable free algebras as Turing-machine undecidability,
cf. e.g. Thm 4.4.3 of \cite{Rep} and \cite{Turing}.}.

We formalise the above discussion into the following definitions:
\begin{definition}
Let $k$ be an infinite field and $A$, a finite dimensional algebra.
\begin{itemize}
\item $A$ is of {\bf finite representation type} if there are only finitely
	many isomorphism classes of indecomposable $A$-modules, otherwise
	it is of infinite type;
\item $A$ is of {\bf tame representation type} if it is of infinite type and
	for any dimension $n$, there is a finite set of 
	$A$-$k[X]$-bimodules\footnote{
		Therefore for the polynomial ring $k[X]$,
		the indeterminate $X$ furnishes the parameter for the one-parameter
		family mentioned in the first paragraph of this subsection. 
		Indeed the indecomposable $k[X]$-modules 
		are classified by powers of irreducible polynomials over $k$.}
	$M_i$ which obey the following:
	\ben
	\item $M_i$ are free as right $k[X]$-modules;
	\item For some $i$ and some indecomposable $k[X]$-module $M$,
		all but a finitely many indecomposable $A$-modules
		of dimension $n$ can be written as $M_i \otimes_{k[X]} M$.
	\een
	If the $M_i$ may be chosen independently of $n$, then we say $A$
	is of {\bf domestic representation type}.
\item $A$ is of {\bf wild representation type} if it is of infinite
	representation type and there is a finitely generated
	$A$-$k[X,Y]$-bimodule $M$ which is free as a right $k[X,Y]$-module
	such that the functor $M \otimes_{k[X,Y]}$ from finite-dimensional
	$k[X,Y]$-modules to finite-dimensional $A$-modules preserves 
	indecomposability and isomorphism classes.
\end{itemize}
\end{definition}
We are naturally led to question ourselves whether the above list is exhaustive.
This is indeed so: what is remarkable is the so-called trichotomy 
theorem which says that all finite dimensional algebras
must fall into one and only one of the above
classification of types\footnote{For a discussion on this theorem and how
	similar structures arises for finite groups, cf. e.g. \cite{Rep,Simson}
	and references therein.}:
\begin{theorem} {\rm (Trichotomy Theorem)}
For $k$ algebraically closed, every finite dimensional algebra $A$ is
of finite, tame or wild representation types, which are mutually exclusive.
\end{theorem}

To this pigeon-hole we may readily apply our path algebras of \sref{ss:quiver}.
Of course such definitions of representation types can be generalised to
additive categories with unique decomposition property. Here by an additive
category $\sf B$ we mean one with finite direct sums and an Abelian
structure on ${\sf B}(X,Y)$, the set of morphisms from object $X$ to $Y$
in $\sf B$ such that the composition map 
${\sf B}(Y,Z) \times {\sf B}(X,Y) \rightarrow {\sf B}(X,Z)$ 
is bilinear for $X,Y,Z$ objects
in $\sf B$. Indeed, that (a) each object in $\sf B$ can be finitely decomposed via
the direct sum into indecomposable objects and that (b) the ring of 
endomorphisms between objects has a unique maximal ideal guarantees that
$\sf B$ possesses unique decomposability as an additive category \cite{Simson}.

The category ${\sf rep}(Q)$, what \cite{Bernstein} calls the {\bf Quiver
Category}, has as its objects the pairs $(V,\alpha)$ with
linear spaces $V$ associated to the nodes and linear mappings $\alpha$, to
the arrows. The morphisms of the category are mappings
$\phi : (V,\alpha) \rightarrow (V',\alpha')$
compatible with $\alpha$ by 
$\phi_{e(l)}\alpha_l = \alpha'_{l}\phi_{s(l)}$.
In the sense of the correspondence between representation of quivers and path 
algebras as discussed in \sref{ss:quiver}, the category ${\sf rep}(Q)$ 
of finite dimensional representations of $Q$,
as an additive category, is equivalent to ${\sf mod}(kQ)$, the category of finite
dimensional (right) modules of the path algebra $kQ$ associated to $Q$.
This equivalence
\[
{\sf rep}(Q) \cong {\sf mod}(kQ)
\]
is the axiomatic statement of the correspondence and justifies why we can hereafter
translate freely between the concept of representation types of quivers
and associated path algebras.

\subsection{Restrictions on the Shapes of Quivers} \label{ss:theorem}
Now we return to our quivers and in particular combine \sref{ss:quiver} and 
\sref{ss:type} to address the problem of how the representation types of the path algebra
restricts the shapes of the quivers. Before doing so let us first justify, as advertised in
\sref{ss:quiver}, why \fref{f:dynkin} and \fref{f:euclid} are canonical.
We first need a preparatory definition: we say a labeled graph $T_1$ is {\bf smaller} 
than $T_2$ if there is an
injective morphism of graphs $\rho : T_1 \rightarrow T_2$ such that for
each edge $(i \stackrel{\gamma}{\rightarrow} j)$ in $T_1$,
$a_{ij} \le a_{\rho(i)\rho(j)}$ (and $T_1$ is said to be strictly
smaller if $\rho$ can not be chosen to be an isomorphism).
With this concept, we can see that the Dynkin and Euclidean graphs 
are indeed our archetypal examples of labeled graphs due to the following theorem:
\begin{theorem} {\rm \cite{Rep,Mod}}
Any connected labeled graph $T$ is one and only one of the 
following:
\ben
	\item $T$ is Dynkin (finite or infinite);
	\item There exists a Euclidean graph smaller than $T$.
\een
\end{theorem}
This is a truly remarkable fact which dictates that the atomic constituents of
all labeled graphs are those arising from semi-simple (ordinary and affine) 
Lie Algebras. The omni-presence of such meta-patterns is still largely
mysterious (see e.g. \cite{He-Song,Gannon} for discussions on this point).

Let us see another manifestation of the elementarity of the Dynkin and Euclidean
Graphs. Again, we need some rudimentary notions.
\begin{definition}
The {\bf Cartan Matrix} for a labeled graph $T$ with labels
$(a_{ij},a_{ji})$ for the edges is the matrix\footnote{This definition
	is inspired by, but should be confused with, Cartan matrices for semisimple
	Lie algebras; to the latter we shall refer as Dynkin-Cartan matrices.
	Also, in the definition we have summed over edges $\gamma$ adjoining
	$i$ and $j$ so as to accommodate multiple edges between the two nodes
	each with non-trivial labels.}
$c_{ij} := 2 \delta_{ij} - \sum\limits_\gamma a^\gamma_{ij}$
\end{definition}
We can symmetrise the Cartan matrix for valued graphs as
$\tilde{c}_{ij} = c_{ij}f_j$ with $\{f_j\}$ the valuation of the nodes
of the labeled graph.
With the Cartan matrix at hand, let us introduce an important 
function on labeled graphs:
\begin{definition} \label{additive}
A {\bf subadditive function} $n(x)$ on a labeled graph $T$ is a function
taking nodes $x \in T$ to $n \in \Q^+$ such that 
$\sum\limits_i n(i) c_{ij} \ge 0~\forall~j$.
A subadditive function is {\bf additive} if the equality holds.
\end{definition}
It turns out that imposing the existence such a function highly restricts
the possible shape of the graph; in fact we are again led back to our 
canonical constituents. This is dictated by the following
\begin{theorem} {\rm (Happel-Preiser-Ringel \cite{Rep})}
\label{thm:HPR}
Let $T$ be a labeled graph and $n(x)$ a subadditive function thereupon,
then the following holds:
\ben
	\item $T$ is either (finite or infinite) Dynkin or Euclidean;
	\item If $n(x)$ is not additive, then $T$ is finite Dynkin or $A_\infty$;
	\item If $n(x)$ is additive, then $T$ is infinite Dynkin or Euclidean;
	\item If $n(x)$ is unbounded then $T = A_\infty$
\een
\end{theorem}
We shall see in the next section what this notion of graph additivity
\cite{Rep,Gannon} signifies for super-Yang-Mills theories. For now,
let us turn to the {\it Theorema Egregium} of Gabriel that definitively
restricts the shape of the quiver diagram once the {\it finitude} of
the representation type of the corresponding path algebra is imposed.
\begin{theorem} {\rm (Gabriel \cite{Gabriel1,Gabriel2,Simson})}
\label{thm:Gab}
A finite quiver $Q$ (and hence its associated path algebra over an
infinite field)
is of finite representation type if and only if it is a disjoint
union of Dynkin graphs of type $A_n$, $D_n$ and $E_{678}$, i.e.,
the ordinary simply-laced $ADE$ Coxeter-Dynkin diagrams.
\end{theorem}
In the language of categories \cite{Bernstein}, where a proof of the
theorem may be obtained
using Coxeter functors in the Quiver Category, the above proposes that
the quiver is (unions of) $ADE$ if and only if there are a 
finite number of non-isomorphic 
indecomposable objects in the category {\sf rep}$(Q)$.

Once again appears the graphs of \fref{f:dynkin}, and in fact only
the single-valence ones: that ubiquitous $ADE$ meta-pattern! We recall 
from discussions in \sref{ss:quiver}
that only for the simply-laced (and thus simply-valanced quivers)
cases, viz. $ADE$ and $\widehat{ADE}$, do the
labels $a^\gamma_{ij}$ precisely prescribe the adjacency matrices.
To what type of path algebras then, one may ask, do the affine 
$\widehat{ADE}$ Euclidean graphs correspond? The answer is given by
Nazarova as an extension to Gabriel's Theorem.
\begin{theorem} {\rm (Nazarova \cite{Poly2,Simson})}
Let $Q$ be a connected quiver without oriented cycles and let
$k$ be an algebraically closed field, then $kQ$ is of
tame (in fact domestic)
representation type if and only if $Q$ is the one of the Euclidean 
graphs of type $\hat{A}_n$, $\hat{D}_n$ and $\hat{E}_{678}$, i.e.,
the affine $ADE$ Coxeter-Dynkin diagrams.
\end{theorem}
Can we push further? What about the remaining quivers of in our canonical
list? Indeed, with the introduction of modulation on the quivers, 
as introduced in \sref{ss:quiver}, the results can be further
relaxed to include more graphs, in fact all the Dynkin and Euclidean
graphs:
\begin{theorem} {\rm (Tits, Bernstein-Gel'fand-Ponomarev,
	Dlab-Ringel, Nazarova-Ringel 
	\cite{Bernstein,Dlab,Poly1,Rep})}
\label{thm:gen}
Let $Q$ be a connected modulated quiver, then
\ben
\item If $Q$ is of finite representation type then $Q$ is Dynkin;
\item If $Q$ is of tame representation type, then $Q$ is Euclidean.
\een
\end{theorem}
This is then our dualism, on the one level of having finite graphs
encoding a (classifiability) infinite algebra and on another level
having the two canonical constituents of all labeled graphs being
partitioned by finitude versus infinitude\footnote{This is much in the
	spirit of that wise adage, ``Cette opposition nouvelle, 
	`le fini et l'infini', ou mieux `l'infini dans le fini',
	remplace le dualisme de l'\^etre et du para\^{\i}tre: ce qui
	para\^{\i}t, en effet, c'est seulement un {\it aspect} de 
	l'objet et l'objet est tout entier {\it dans} 
	cet aspect et tout entier hors de lui \cite{Sartre}.''}.

\section{Quivers in String Theory and Yang-Mills in Graph Theory} \label{s:marriage}
We are now equipped with a small arsenal of facts; it is now our duty to 
expound upon them. Therefrom we shall witness how
axiomatic studies of graphs and representations may shed light on current
developments in string theory.

Let us begin then, upon examining condition (\ref{orb_cond}) and Definition 
\ref{additive}, with the following

\begin{observation}
\label{ob:fin}
The condition for finitude of ${\cal N} = 2$ orbifold SYM theory is equivalent to
the introduction of an additive function on the corresponding quiver as a
labeled graph.
\end{observation}

This condition that for the label $n_i$ to each node $i$ and adjacency matrix
$A_{ij}$, $2 n_i = \sum\limits_j a_{ij} n_j$ is a very
interesting constraint to which we
shall return shortly. What we shall use now is Part 3 of 
Theorem \ref{thm:HPR} in conjunction with the above observation to deduce

\begin{corollary} \label{cor:N2}
All finite ${\cal N} = 2$ super-Yang-Mills Theories with bi-fundamental matter
have their quivers as (finite disjoint unions) of the single-valence (i.e.,
$(1,1)$-labeled edges) cases
of the Euclidean (\fref{f:euclid}) or Infinite Dynkin (\fref{f:dynkin}) graphs.
\end{corollary}

A few points to remark. This is slightly a more extended list than that given in
\cite{Mirror} which is comprised solely of the $\widehat{ADE}$ quivers.
These latter cases are the ones of contemporary interest because they,
in addition to being geometrically constructable (Cf. \sref{ss:geo}),
are also obtainable from the string orbifold technique\footnote{
	And in the cases of $A$ and $D$ also from Hanany-Witten
	setups \cite{Kapustin,ZD}.} (Cf. \sref{ss:probe})
since after all the finite discrete subgroups of $SU(2)$ fall into
an $\widehat{ADE}$ classification due to McKay's Correspondence 
\cite{McKay,Orb1,Han-He}.
In addition to the above well-behaved cases, we also 
have the infinite simply-laced Dynkin graphs: $A_\infty,D_\infty$ and
$A_\infty^\infty$. The usage of the Perron-Frobenius Theorem in
\cite{Mirror} restricts one's attention to finite matrices.
The allowance for infinite graphs of course implies an infinitude of nodes
and hence infinite products for the gauge group. One needs not exclude
these possibilities as after all in the study of D-brane probes,
Maldacena's large $N$ limit has been argued in \cite{KS,LNV,Han-S}
to be required for conformality and finiteness. In this limit of
an infinite stack of D-branes, infinite gauge groups may well arise.
In the Hanany-Witten picture, $A_\infty^\infty$ for example would 
correspond to an infinite array of NS5-branes, and $A_\infty$,
a semi-infinite array with enough D-branes on the other side
to ensure the overall non-bending and parallelism of the NS.
Such cases had been considered in \cite{Erlich}.

Another comment is on what had been advertised earlier in \sref{ss:quiver}
regarding the adjacency matrices. Theorem \ref{thm:HPR} does not exclude
graphs with multiple-valanced oriented labels. This issue does not
arise in ${\cal N}=2$ which has only single-valanced and unoriented
quivers.
However, going beyond to ${\cal N} = 1,0$, requires generically oriented and 
multiply-valanced quivers (i.e., non-symmetric, non-binary matter matrices)
\cite{Han-He,Su4}; or, it is conceivable that certain theories
not arising from orbifold procedures may also possess these generic traits.
Under this light we question ourselves how one may
identify the bi-fundamental matter matrices not with strict adjacency
matrices of graphs but with the graph-label matrices $a_{ij}^\gamma$
of \sref{ss:quiver} so as to accommodate multiple, chiral bi-fundamentals
(i.e. multi-valence, directed graphs). In other words, could 
Corollary \ref{cor:N2} {\it actually be relaxed to incorporate all of 
the Euclidean and infinite Dynkin graphs} as dictated by Theorem \ref{thm:HPR}?
Thoughts on this direction, viz., how to realise Hanany-Witten brane
configurations for non-simply-laced groups have been engaged but still
waits further clarification \cite{chat}.

Let us now turn to Gabriel's famous Theorem \ref{thm:Gab} and see its implications
in string theory and vice versa what information the latter provides for
graph theory. First we make a companion statement to Observation \ref{ob:fin}:

\begin{observation}
\label{ob:af}
The condition for asymptotically free ($\beta < 0$) ${\cal N} = 2$
SYM theory with bi-fundamentals
is equivalent to imposing a subadditive (but not additive) function
of the corresponding quiver.
\end{observation}

This may thus promptly be utilised together with Part 2 of Theorem \ref{thm:HPR} to
conclude that the only such theories are ones with $ADE$ quiver, or,
allowing infinite gauge groups, $A_\infty$ as well (and indeed all finite
Dynkin quivers once, as mentioned above, non-simply-laced groups have been resolved).
This is once again a slightly extended version of the results in \cite{Mirror}.

Let us digress, before trudging on, a moment to consider what is means to
encode SYM with quivers. Now we recall that for the quiver $Q$, 
the assignment of objects and morphisms to the category {\sf rep}$(Q)$,
or vector spaces and linear maps to nodes $Q_0$ and edges $Q_1$ in $Q$,
or bases to the path algebra $kQ$, are all equivalent procedures.
From the physics perspective, these assignments are precisely what
we do when we associate vector multiplets to nodes and hypermultiplets to
arrows as in the orbifold technique, or NS-branes to nodes and oriented
open strings between D-branes to arrows as in the Hanany-Witten configurations, or 
singularities in Calabi-Yau to nodes and colliding fibres to arrows as in
geometrical engineering. In other words the three methods, \sref{ss:quiver},
\sref{ss:probe} and \sref{ss:geo}, of constructing
gauge theories in four dimensions currently in vogue are different
representations of {\sf rep}$(Q)$ and are hence {\it axiomatically}
equivalent as far as quiver theories are concerned.

Bearing this in mind, and in conjunction with Observations \ref{ob:fin}
and \ref{ob:af}, as well as Theorem \ref{thm:Gab} together with its
generalisations, and in particular Theorem \ref{thm:gen}, we make the following

\begin{corollary} \label{cor:alg}
To an asymptotically free ${\cal N}= 2$ SYM with bi-fundamentals is associated a
finite path algebra and to a finite one, a tame path algebra. The association
is in the sense that these SYM theories (or some theory categorically equivalent thereto)
prescribe representations of the only quivers of such representation types.
\end{corollary}

What is even more remarkable perhaps is that due to the Trichotomy Theorem,
the path algebra associated to
{\bf all other quivers} must be of wild representation type. What this means,
as we recall the unclassifiability of algebras of wild representations, is
that these quivers are unclassifiable. In particular, if we assume that
SYM with ${\cal N}=0,1$ and arbitrary bi-fundamental matter content can
be constructed (either from orbifold techniques, Hanany-Witten, or geometrical 
engineering), then these theories {\it can not} be classified, in the strict
sense that they are Turing undecidable and there does not exist, in any finite
language, a finite scheme by which they could be listed. Since the set of SYM with
bi-fundamentals is a proper subset of all SYM, the like applies to general SYM.
What this signifies is that however ardently we may continue to provide
more examples of say finite ${\cal N}=1,0$ SYM, the list can never be finished
nor be described, unlike the ${\cal N}=2$ case where the above discussions
exhaust their classification. We summarise this amusing if not depressing
fact as follows:

\begin{corollary} \label{cor:un}
The generic ${\cal N} = 1,0$ SYM in four dimensions are unclassifiable 
in the sense of being Turing undecidable.
\end{corollary}

We emphasise again that by unclassifiable here we mean not {\it
completely} classifiable because we have given a subcategory (the
theories with bi-fundamentals) which is unclassifiable. Also, we rest upon
the assumption that for any bi-fundamental matter content an SYM could
be constructed. Works in the direction of classifying all possible
gauge invariant operators in an ${\cal N}=1$ SUSY Lagrangian have been
pursued \cite{Skenderis}. Our claim is much milder as no further
constraints than the possible na\"{\i}ve matter content are imposed; we
simply state that the complete generic problem of classifying the
${\cal N}<2$ matter content is untractable. In \cite{Skenderis}, the
problem has been reduced to manipulating a certain cohomological
algebra; it would be interesting to see for example, whether such BRST
techniques may be utilised in the classification of certain categories
of graphs.

Such an infinitude of gauge theories need not worry us as there certainly
is no shortage of say, Calabi-Yau threefolds which may be used to
geometrically engineer them.
This unclassifiability is rather in the spirit of that of, for example,
four-manifolds.
Indeed, though we may never exhaust the list, we are not precluded
from giving large exemplary subclasses which are themselves
classifiable, e.g., those prescribed by the orbifold theories. 
Determining these theories amounts to the classification of the finite
discrete subgroups of $SU(n)$. 

We recall from Corollary \ref{cor:N2} that
${\cal N}=2$ is given by the affine and infinite Coxeter-Dynkin graphs of
which the orbifold theories provide the $\widehat{ADE}$ cases. What
remarks could one make for ${\cal N} = 0,1$, i.e., $SU(3,4)$ 
McKay quivers \cite{Han-He,Su4}?
Let us first see ${\cal N}=2$ from the graph-theoretic perspective, which
will induce a relationship between additivity (Theorem \ref{thm:HPR}) and
Gabriel-Nazarova (Theorems \ref{thm:Gab} and extensions).
The crucial step in Tit's proof of Gabriel's Theorem is the introduction
of the quadratic form on a graph \cite{Bernstein,Donovan}:
\begin{definition}
For a labeled quiver $Q = (Q_0,Q_1)$, one defines the (symmetric bilinear)
{\bf quadratic form} $B(x)$ on the set $x$ of the labels as follows:
\[
B(x) := \sum\limits_{i \in Q_0} x_i^2 - \sum\limits_{\alpha \in Q_1}
	x_{s(\alpha)} x_{e(\alpha)}.
\]
\end{definition}
The subsequent work was then to show that finitude of representation is
equivalent to the positive-definity of $B(x)$, and in fact, as in
Nazarova's extension, that tameness is equivalent to positive-semi-definity.
In other words, finite or tame representation type can be translated,
in this context, to a Diophantine inequality which dictates the nodes and
connectivity of the quiver (incidentally the very same Inequality which
dictates the shapes of the Coxeter-Dynkin Diagrams or the vertices and
faces of the Platonic solids in $\R^3$):
\[
B(x) \ge 0 \Leftrightarrow \widehat{ADE},ADE~~~~~~~B(x) > 0 \Leftrightarrow ADE.
\]
Now we note that $B(x)$ can be written as 
$\frac 12 x^T \cdot c \cdot x$ where $(c)_{ij}$ is {\it de facto} 
the Cartan Matrix for graphs as
defined in \sref{ss:theorem}. The classification problem thus, because
$c := 2I - a$, becomes
that of classifying graphs whose adjacency matrix $a$ has maximal eigenvalue
2, or what McKay calls $C_2$-graphs in \cite{McKay}. This issue was addressed
in \cite{Smith} and indeed the $\widehat{ADE}$ graphs emerge.
Furthermore the additivity condition $\sum\limits_j c_{ij} x_j \ge 0~\forall~i$
clearly implies the constraint $\sum\limits_{ij} c_{ij} x_i x_j \ge 0$ 
(since all labels are positive) and thereby the like on the quadratic form.
Hence we see how to arrive at the vital step in Gabriel-Nazarova through
graph subadditivity.

The above discussions relied upon the specialty of the number 2.
Indeed one could translate between the graph quadratic form $B(x)$
and the graph Cartan matrix precisely because the latter is defined by
$2I - a$. From a physical perspective this is precisely the discriminant function
for ${\cal N}=2$ orbifold SYM (i.e. $d = 2$) as discussed at the
end of \sref{ss:probe}. This is why $\widehat{ADE}$ arises in all these contexts.
We are naturally led to question ourselves, what about general\footnote{
	In the arena of orbifold SYM, $d=1,2,3$, but in a broader settings,
	as in generalisation of McKay's Correspondence, $d$ could be
	any natural number.} $d$?
This compels us to consider a {\bf generalised Cartan matrix} for graphs
(Cf. Definition in \sref{ss:theorem}), given
by $c_{ij} := d \delta_{ij} - a_{ij}$, our discriminant function of \sref{ss:probe}.
Indeed such a matrix was considered in \cite{Steinberg} for general McKay quivers.
As a side remarks, due to such an extension, Theorem \ref{thm:HPR} must likewise
be adjusted to accommodate more graphs; a recent paper \cite{McKay2} shows
an example, the so-dubbed semi-Affine Dynkin Diagrams, where a new class of
labeled graphs with additivity with respect to the extended $c_{ij}$ emerge.

Returning to the generalised Cartan matrix, in \cite{Steinberg},
the McKay matrices $a_{ij}$ were obtained, for an arbitrary
finite group $G$, by tensoring a faithful $d$-dimensional representation with
the set of irreps: $r_d \otimes r_i = \oplus_j a_{ij} r_j$.
What was noticed was that the scalar product defined with respect
to the matrix $d \delta_{ij} - a_{ij}$ (precisely our generalised Cartan) was
positive semi-definite in the vector space $V = \{x_i\}$ of labels.
In other words, $\sum\limits_{ij} c_{ij} x_i x_j \ge 0$.
We briefly transcribe his proof in the Appendix.
What this means for us is that is the following

\begin{corollary}
\label{cor:d}
String orbifold theories can not produce a completely IR free 
(i.e., with respect to all semisimple components of the gauge group) 
QFT (i.e., Type (1), $\beta > 0$).
\end{corollary}

To see this suppose there existed such a theory. Then $\beta > 0$, implying
for our discriminant function that $\sum\limits_j c_{ij} x_j < 0~\forall~i$ for
some finite group.
This would then imply, since all labels are positive, that 
$\sum\limits_{ij} c_{ij} x_i x_j < 0$, violating the positive semidefinity
condition that it should always be nonnegative for any finite group according to
\cite{Steinberg}. Therefore by {\it reductio ad absurdum}, we conclude 
Corollary \ref{cor:d}.

On a more general setting, if we were to consider using the
generalised Cartan matrix $d \delta_{ij} - a_{ij}$
to define a generalised subadditive function (as opposed to merely $d=2$),
could we perhaps have an extended classification scheme?
To our knowledge this is so far an unsolved problem for indeed take
the subset of these graphs with all labels being 1 and 
$d n_i = \sum\limits_j a_{ij} n_j$, these are known as $d$-regular graphs
(the only 2-regular one is the $\widehat{A}$-series) and these are
already unclassified for $d>2$. We await input from mathematicians on this point.

\section{Concluding Remarks and Prospects}
The approach of this writing has been bilateral. On the one hand, we
have briefly reviewed the three contemporary techniques of obtaining 
four dimensional gauge theories from string theory, namely
Hanany-Witten, D-brane probes and geometrical engineering. In
particular, we focus on what finitude signifies for these theories and
how interests in quiver diagrams arises. Subsequently, we approach
from the mathematical direction and have taken a promenade in the
field of axiomatic representation theory of algebras associated to
quivers. The common ground rests upon the language of graph theory,
some results from which we have used to address certain issues in 
string theory.

From the expression of the one-loop $\beta$-function, we have defined
a discriminant function $f := d \delta_{ij} - a_{ij}^d$ for the quiver with
adjacency matrix $a_{ij}$ which encodes the bi-fundamental matter
content of the gauge theory. The nullity (resp. negativity/positivity)
of this function gives a necessary condition for the finitude
(resp. IR freedom/asymptotic freedom) of the associated gauge theory.
We recognise this function to be precisely the generalised Cartan
matrix of a (not necessarily finite) graph and the nullity (resp.
negativity) thereof, the additivity (resp. strict subadditivity) of
the graph. In the case of $d=2$, such graphs are completely
classified: infinite Dynkin or Euclidean if $f = 0$ and finite Dynkin
or $A_\infty$ if $f < 0$. In physical terms, this means that these
are the {\it only} ${\cal N}=2$ theories with bi-fundamental matter
(Corollary \ref{cor:N2} and Observation \ref{ob:af}).
This slightly generalises the results of \cite{Mirror} by the
inclusion of infinite graphs, i.e., theories with infinite product
gauge groups. From the mathematics alone, also included are the
non-simply-laced diagrams, however we still await progress in the
physics to clarify how these gauge theories may be fabricated.

For $d>2$, the mathematical problem of their classification is so far
unsolved. A subclass of these, namely the orbifold theories coming
from discrete subgroups of $SU(n)$ have been addressed upto $n=4$
\cite{Orb2,Han-He,Su4}. A general remark we can make about these
theories is that, due to a theorem of Steinberg, D-brane probes on
orbifolds can never produce a completely IR free QFT (Corollary
\ref{cor:d}).

From a more axiomatic stand, we have also investigated possible
finite quivers that may arise. In particular we have reviewed the
correspondence between a quiver and its associated path algebra. Using
the Trichotomy theorem of representation theory, that all finite
dimensional algebras over an algebraically closed field are of either
finite, tame or wild type, we have seen that all quivers are
respectively either $ADE$, $\widehat{ADE}$ or unclassifiable. In
physical terms, this means that asymptotically free and finite 
${\cal N}=2$ SYM in four dimensions respectively exhaust the {\it
only} quiver theories of respectively finite and tame type (Corollary
\ref{cor:alg}). What these particular path algebras mean in a physical
context however, is yet to be ascertained. For the last type, we have
drawn a melancholy note that all other theories, and in particular,
${\cal N}<2$ in four dimensions, are in general Turing
unclassifiable (Corollary \ref{cor:un}).

Much work remains to be accomplished. It is the main purpose of this
note, through the eyes of a neophyte, to inform readers in each of two
hitherto disparate fields of gauge theories and axiomatic
representations, of certain results from the other. It is hoped that
future activity may be prompted.

\section*{Acknowledgements}
{\it Ad Catharinae Sanctae Alexandriae et Ad Majorem Dei Gloriam...\\}
I extend my sincere gratitude to A. Hanany for
countless valuable suggestions and comments on the paper. Also I
would like to acknowledge B. Feng for extolling the virtues of branes,
L. Ng, those of n-regular graphs and J. S. Song, those of geometrical
engineering, as well as the tireless discussions these my
friends and collegues have afforded me.
Moreover, I am indebted to K. Skenderis for helpful discussions and
his bringing \cite{Skenderis} to my attention,
N. Moeller for reference \cite{Sartre} and N. Tkachuk
for offering to translate the extensive list of Russian documents in
which are buried priceless jewels.
Furthermore, I am thankful to the organisers of the
``Modular Invariants, Operator Algebras and Quotient Singularities
Workshop'' and the Mathematics Research Institute of the University of
Warwick at Conventry, U. K. for providing the friendly atmosphere
wherein much fruitful discussions concerning the unification through
ADE were engaged.

Indeed, I am much obliged to M. Warden for charming inspirations,
the CTP and the NSF for their gracious patronage as well as the ITP
for her warm hospitality.

\section*{Appendix}
We here transcribe Steinberg's proof of the semi-definity of the scalar product
with respect to the generalised Cartan matrix, in the vector space
$V = \{x_i \in \Z_+ \}$ of labels \cite{Steinberg}.
Our starting point is (\ref{aij}), which we re-write here as
\[
	r_d \otimes r_i = \bigoplus\limits_{j} a_{ij} r_j
\]
First we note that, if $\bar{i}$ is the dual representation to $i$,
then $a_{ij} = a_{\bar{j}\bar{i}}$ by taking the conjugates (dual)
of both sides of (\ref{aij}). Whence we have
\begin{lemma}
\label{app:dimaij} For $d_i = \dim{r_i}$,
	$d d_i = \sum\limits_j a_{ij} d_j = \sum\limits_j a_{ji} d_j$.
\end{lemma}
The first equality is obtained directly by taking the dimension of
both sides of (\ref{aij}) as in (\ref{dimaij}). To see the second
we have $d d_i = d d_{\bar{i}}$ (as dual representations have the
same dimension) which is thus equal to $\sum\limits_j a_{\bar{i}j} d_j$,
and then by the dual property $a_{ij} = a_{\bar{j}\bar{i}}$ above
becomes $\sum\limits_{\bar{j}} a_{\bar{j}i} d_{\bar{j}} =
\sum\limits_j a_{ji} d_j$. QED.

Now consider the following for the scalar product:
\begin{eqnarray*}
2 \sum_{ij} c_{ij} x_i x_j & = & 2 \sum_{ij} (d \delta_{ij} - a_{ij}) x_i x_j
		= 2 (d \sum_i x_i^2 - \sum_{ij} a_{ij} x_i x_j)	\\
	& = & 2 (\sum_i (d - a_{ii}) x_i^2 - \sum_{i \ne j} a_{ij} x_i x_j) \\
	& = & 2 \sum_i \frac{1}{2} (\frac{1}{d_i} \sum_j a_{ij} d_j +
		\frac{1}{d_i} \sum_j a_{ji} d_j - a_{ii}) x_i^2 -
		\sum_{i \ne j} a_{ij} x_i x_j)~~\mbox{(by Lemma \ref{app:dimaij})} \\
	& = & \sum_{i \ne j} (a_{ij} + a_{ji}) \frac{d_j}{d_i} x_i^2 - 2 a_{ij} x_i x_j
		= \sum_{i < j} (a_{ij} + a_{ji}) (\frac{d_j}{d_i} x_i^2 
		+ \frac{d_i}{d_j} x_j^2- 2 x_i x_j) \\
	& = & \sum_{i < j} (a_{ij} + a_{ji}) \frac{(d_j x_i - d_i x_j)^2}{d_i d_j} \ge 0
\end{eqnarray*}
From which we conclude
\begin{proposition} {\rm (Steinberg)}
	In the vector space of positive labels, the scalar product is positive 
	semi-definite, i.e., $\sum\limits_{ij} c_{ij} x_i x_j \ge 0$.
\end{proposition}


\end{document}